\documentclass[superscriptaddress,twocolumn,aps,prb,floatfix]{revtex4}
\usepackage{graphicx}
\usepackage{amssymb,amsmath}
\usepackage{array}
\usepackage{dcolumn}
\usepackage{multirow}
\usepackage{color}      % use if color is used in text
\usepackage{subfigure} % use for side-by-side figures

\usepackage[french,english]{babel}
\usepackage[utf8]{inputenc}

\begin{document}
 
%========================================================================= 
\title{First principles calculation of the phonons modes in the hexagonal $\rm
YMnO_3$ ferroelectric and paraelectric phases}
%=========================================================================
 
\author{Julien Varignon}
\author{S\'ebastien Petit}
\author{Marie-Bernadette Lepetit}
 
\affiliation{CRISMAT, ENSICAEN-CNRS UMR6508, 6~bd. Mar\'echal Juin, 14050 Caen, 
FRANCE}

%=========================================================================
\date{\today}
%=========================================================================

\begin{abstract}
  The lattice dynamics of the $\rm YMnO_3$ magneto-electric compound has been
  investigated using density functional calculations, both in the
  ferroelectric and the paraelectric phases. The coherence between the
  computed and experimental data is very good in the low temperature
  phase. Using group theory, modes continuity and our calculations we were
  able to show that the phonons modes observed by Raman scattering at 1200K
  are only compatible with the ferroelectric $P6_{3} cm$ space group, thus
  supporting the idea of a ferroelectric to paraelectric phase transition at
  higher temperature. Finally we proposed a candidate for the phonon part of
  the observed electro-magnon. This mode, inactive both in Raman scattering
  and in Infra-Red, was shown to strongly couple to the Mn-Mn magnetic
  interactions.

\pacs{63.20.-e,63.20.dk,63.20.kd}
\end{abstract}

%=========================================================================
\maketitle                                                                            

%=========================================================================
\section{Introduction} \label{intro}
%=========================================================================
Materials presenting magneto-electric coupling exhibit 
magnetic properties  coupled to the electric properties, such as
polarization or dielectric constant. These materials have attracted a lot of
attention over the last years since the magneto-electric coupling allows a
possible control of the magnetic properties by an electric field and 
over electric properties using a magnetic field.

Unfortunately, the microscopic origin of the coupling between the magnetic and
electric order parameters is still ill known. The knowledge of phonons spectra
can however bring help on understanding this coupling. Indeed, not only the
phonons modes are strongly related to the existence and amplitude of a
spontaneous polarization, but in addition strong spin-phonons coupling occur
in multiferroic material. This coupling can even be strong enough in order to
result in hybrid excitations built from the mixing between phonons and
spin-waves~\cite{EM}. The existence of such hybrid modes, called
electromagnons, were recently discovered in $\rm RMnO_3$ orthorhombic
manganites~\cite{EM-RMnO3} using optical measurements. More recently such excitations
were also found by inelastic neutrons scattering in the hexagonal manganite $\rm
YMnO_3$~\cite{Petit07,Pailhes09}.

\begin{figure}[h]
\begin{minipage}[t]{3ex}
(a)
\end{minipage}
\begin{minipage}[c]{3.2cm}
\resizebox{3.2cm}{!}{\includegraphics{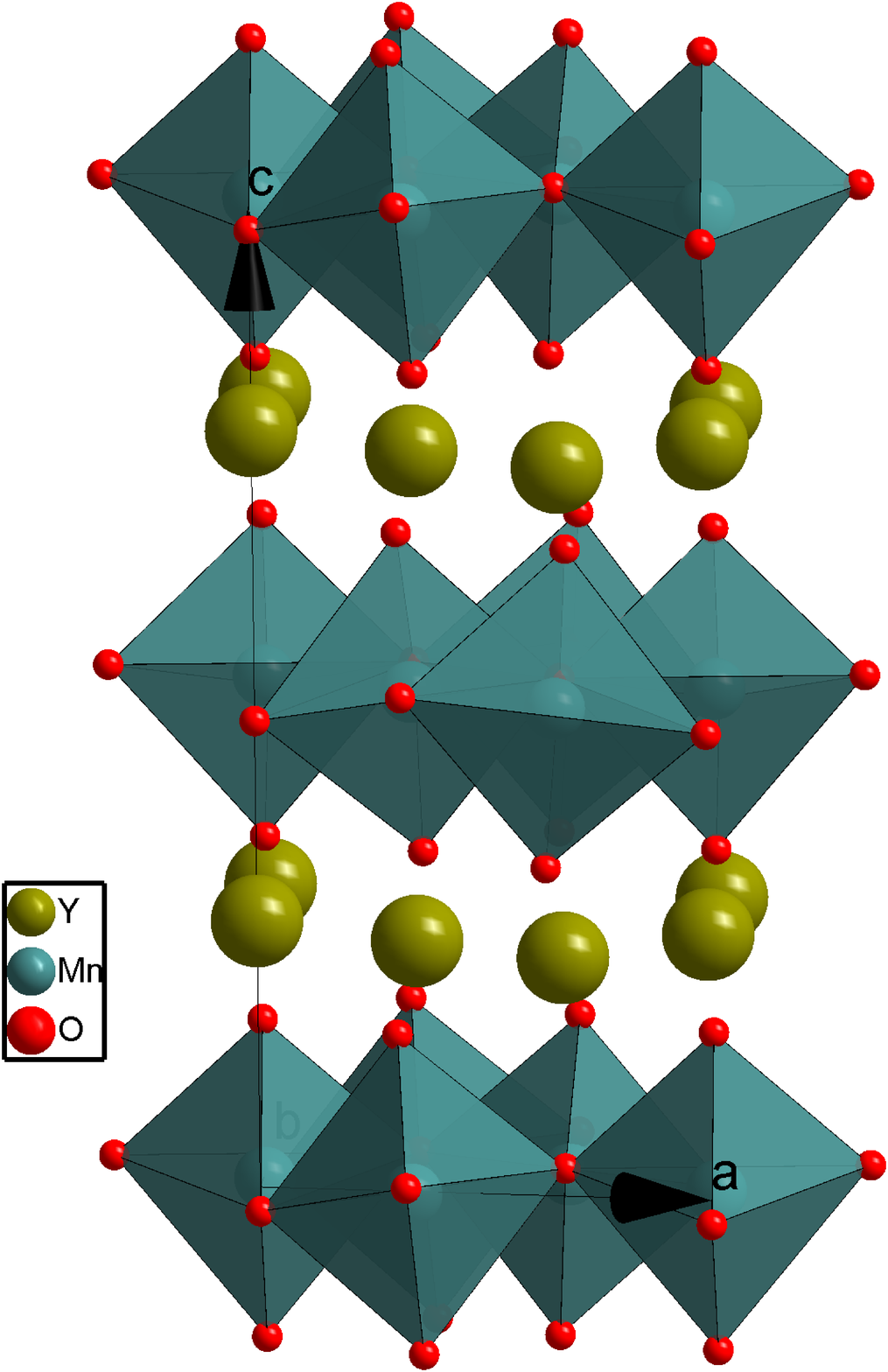}}
\end{minipage}
\begin{minipage}[t]{3ex}
(b)
\end{minipage}
\begin{minipage}[c]{4.2cm}
\resizebox{4.5cm}{!}{\includegraphics[angle=0]{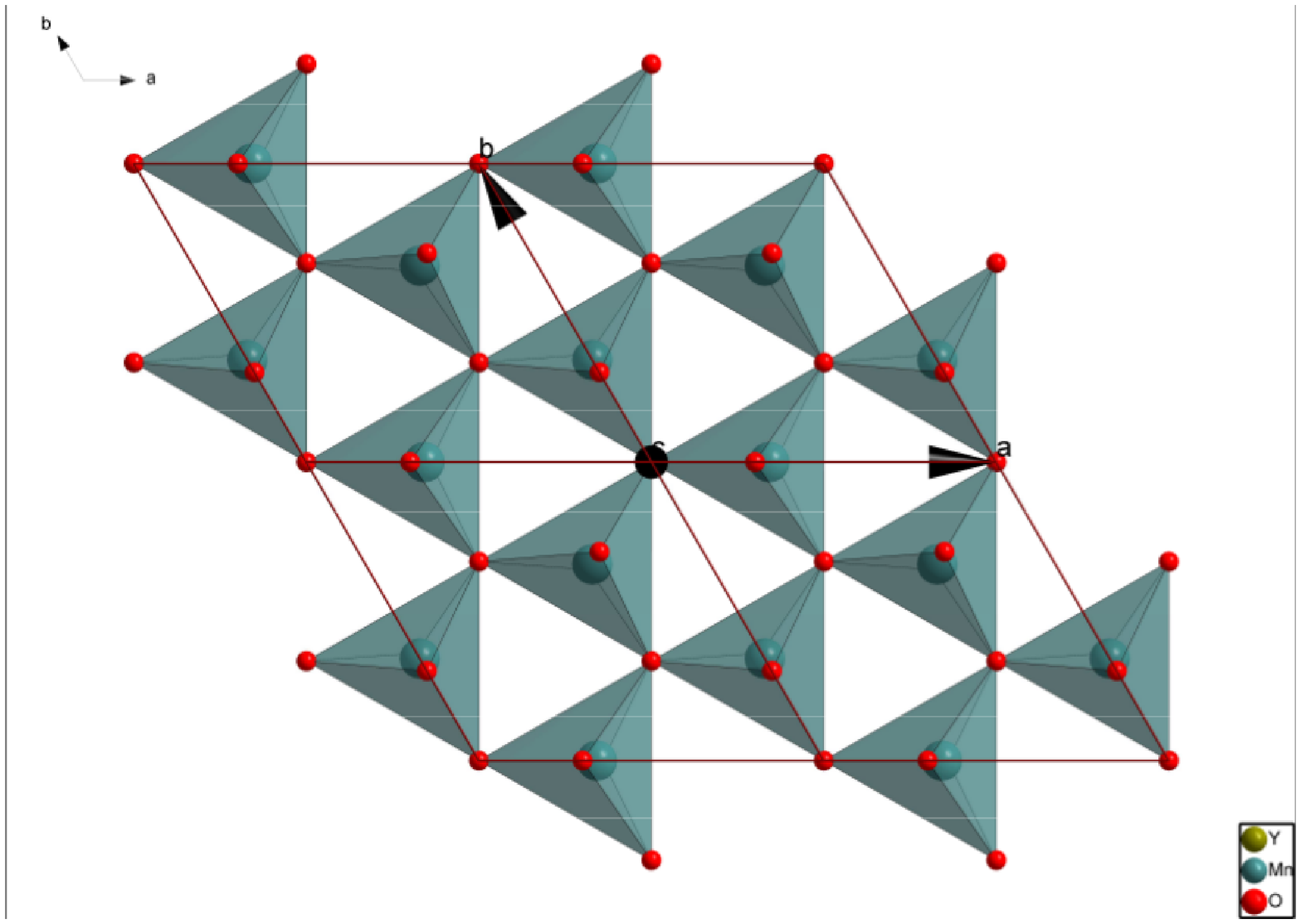}}
\end{minipage}
\caption{(a) representation of the $\rm YMnO_3$ crystal structure. (b) view
  along the $\vec c$ direction of the $\rm MnO_5$ bipyramids layers. }
\label{fig:struc}
\end{figure}

Hexagonal $\rm YMnO_3$ is a layered compound where the manganese ions are
located in triangle-based bipyramids. These bipyramids are arranged in 
planes parallel to the $(\vec a,\vec b)$ direction so that the manganese ions
form a distorted triangular lattice (see figure~\ref{fig:struc}). They share
an oxygen atom in the $(\vec a,\vec b)$ planes. The yttrium atom is located in
between the bipyramids layers.
$\rm YMnO_3$ is paraelectric at high temperature with space group $P6_3/m
mc$. Under a critical temperature it is ferroelectric, with space group
$P6_{3} cm$~; the polarization is aligned along the $\vec c$ direction. The
temperature of this paraelectric (PE) to ferroelectric (FE) phase transition
is however still under debate. Indeed, usually assumed to be around $\simeq
920\,K$~\cite{TE=920K}, it was recently proposed to occur at a higher
temperature, $\simeq 1050-1100\,K$~\cite{TE=1100K,TE-Abra}. A very recent
neutrons diffraction investigation~\cite{TE-Gibbs} suggests the existence of
an iso-symmetric phase transition at $\simeq 920\,K$, resulting in a strong
lowering of the polarization amplitude, while the non-centrosymmetric to
centrosymmetric transition occurs at $1258\pm4\,K$ only. At much lower
temperature ($T_N\simeq 75\,K$) $\rm YMnO_3$ undergoes a paramagnetic (PM) to
antiferromagnetic (AFM) transition. The $\rm Mn^{3+}$ ions are in a high spin
state $S=2$. Neutrons diffraction experiments witness an in-plane spins
orientation associated with a $120^\circ$ ordering, with the spins
perpendicular to the Mn--O bonds~\cite{spins,TN-sp}. The ferroelectric and
antiferromagnetic order parameters are coupled and a giant atomic motion was
revealed at the N\'eel temperature~\cite{Park08}. The magnetic transition is
however an iso-structural transition and the space group remains $P6_{3} cm$ in
the AFM phase~\cite{TN-sp}. The magnetic space group is more subject to
caution. Indeed, while most authors favor an identical $P6_{3} cm$ magnetic
and structural space group~\cite{spins,TN-sp}, the absence of a linear
magneto-electric coupling and the existence of a ferromagnetic component
recently observed~\cite{Pailhes09,YMnO3_1} are incompatible with the $P6_{3} cm$
magnetic group. Group analysis shows that these experimental data can only be
accounted for in the $P6'_3$ magnetic group~\cite{YMnO3_1}.

Phonons spectra are often used to help understanding both the
ferroelectric/paraelectric transitions and the magneto-electric
coupling. While the $\rm YMnO_3$ phonons spectrum has been experimentally
studied both by Raman~\cite{Rm_Iliev,Rm_Fuku,Rm_Vermette} and
Infra-Red~\cite{IR-Vinh} measurements, there is still many opened questions
concerning not only modes assignations but also the fact that several of the
symmetry allowed modes are not experimentally observed. First principle
calculations not only could help understanding the experimental observation
but also could yield more light on the recently observed
electromagnon~\cite{Petit07,Pailhes09}. %However, to our knowledge, no such
%calculations were done up to now. 
The present paper thus presents first principle calculations of the $\rm
YMnO_3$ phonons spectrum in the different space groups associated with the FE
and PE phases, that is $P6_{3} cm$ and $P6_3/m mc$. The next section will be
devoted to the methodological details and preliminary results (geometry
optimization, Born effective charges, polarization, dielectric tensor, etc...)
and section III will describe and discuss phonons results.

\section{Method and preliminary results}
We performed geometry optimization, spontaneous polarization and phonons
frequencies calculations within the Density Functional Theory (DFT) as
implemented in the Crystal09 package~\cite{CRYSTAL09}. Since the manganese 3d
shells are strongly correlated we used hybrid functionals in order to better
take into account the self-interaction cancellation.  Three different
functionals were used in this work, namely the classical B3LYP
functional~\cite{B3LYP}, the recently developed B1WC~\cite{B1WC} functional
for ferroelectric systems and a B1PW~\cite{B1WC} functional corresponding to
the B1WC one with the Perdew-Wang exchange
functional~\cite{Perdrew_PRB33}. Only the B1PW results are presented in the
paper, the results obtained with the other functionals can be found in the
supplementary data. Small core pseudo-potentials were used for the heavy atoms
(Mn and Y) associated with semi-valence and valence $2\zeta$ and $2\zeta$ plus
polarization basis sets~\cite{Bases}. The oxygen ions are represented in an
all-electrons basis set of $2\zeta$ quality specifically optimized for $\rm
O^{2-}$ ions~\cite{Bases}.
% TOLDEE = 10-9
% XLGRID
% 6 6 3 = 14 points K
The mono-determinantal character of the Kohn-Sham approximation does not allow
the possibility to deal with magnetic frustration as observed in $\rm
YMnO_3$. Indeed, only collinear magnetism is possible in the CRYSTAL code.  A
non-magnetic electronic density can be calculated, however it does not account
for the open-shell and high spin character of the $\rm Mn^{3+}$ ions, which
are responsible for a large stabilization energy (of the order of two Hubbard
$3d$-shell $U$ per Mn atom, this is of the order of 10eV) and crucial for a
reasonable electronic structure representation.  A ``{\it pseudo}''
antiferromagnetic configuration can be proposed on a $2\times 2 \times 1$
cell.  In this case, only two among the three antiferromagnetic (AFM)
couplings are satisfied within each triangle. It results in a lower energy for
the antiferromagnetic bonds and an associated loss of symmetry. When the
geometry optimization is performed, the Mn-Mn distances associated with
anti-aligned spins are shorter than the bonds with aligned spins, resulting in
a structure with unphysical symmetry and characteristics. The choice of a
ferromagnetic (FM) configuration for the $\rm Mn^{3+}$ S=2 spins thus seems the
best compromise. Indeed, not only this configuration respects the system
symmetry (a crucial aspect for accurate phonons modes), but also the energy
error that can be associated with the loss of the inter-atomic exchange energy
(that is $1/2 J\simeq 1.5\,\rm meV$ per bond) is much weaker than the error
associated with the loss of the magnetic character for the $\rm Mn^{3+}$
ions. For these reasons all our calculations will be done within a
ferromagnetic Mn spins alignment.

Within this procedure, we computed the polarization (using Berry phase
approach) for the different experimental geometries available in the
literature at different temperatures.  Table~\ref{t:pol} reports the computed
polarization values.  One sees immediately that our computed values are in
very good agreement with the experimental ones, despite the fact that the
magnetic order used in the calculation is ferromagnetic and not
antiferromagnetic. The quality of these results thus validate our choices for
the calculations. In addition it supports the idea, first proposed by Lee {\it
  et al}~\cite{Park08}, that in this system the magneto-electric coupling is
mainly driven by elastic effects rather than by a direct coupling between the
polarization and the magnetic order.

\begin{table}[h!]
\begin{center}
\begin{tabular}{c|ccc}
\hline
Structures & 10\,K~\cite{Park08}  & 180\,K~\cite{VAken} & 300~K\,\cite{Park08}
\\ \hline
Exp. $\rm P_S$ & $\sim$ 1~\cite{YMnO3_1} & 6.2~\cite{VAkenPol} & 4.5~\cite{Pol300} \\
This work & 1.1                & 6.2                 & 4.9  \\
\hline
\end{tabular}
\caption{Spontaneous polarization $\rm P_{\!S}$ in $\mu C.cm^{-2}$ for different
  temperatures. Experimental and computed values using the experimental
  geometries at the given temperatures.}
\label{t:pol}
\end{center}
\end{table}

In a second step we optimized the geometry. Both the cell parameters and the
atomic positions compare very well with the experimental values in the
paramagnetic phase~\cite{VAken}. Indeed, the error on the lattice parameters
is weaker than 0.5\% and for the atomic positions $\sum_{\rm at} |\vec r_{\rm
  at}^{\,\rm exp} - \vec r_{\rm at}^{\,\rm calc}|^2 = 0.0087\rm\,\AA^2$.

\begin{table}[h!]
\begin{center}
\begin{tabular}{cc|ccc|cc}
\hline
\multicolumn{2}{c|}{\multirow{2}{*}{$Z^*$}} & \multirow{2}{*}{x} & \multirow{2}{*}{y} & \multirow{2}{*}{z} & \multicolumn{2}{c}{Dynamic charge}\\
 &&&& & Exp.~[\onlinecite{IR-Vinh}] & This work \\
\hline
\multirow{3}{*}{Y$_1$} & x & 3.5 & 0.0 & 0.0  && \\
                       & y & 0.0 & 3.5 & 0.0  &  4.0 & 3.6\\
                       & z & 0.0 & 0.0 & 3.9 && \\
\hline
\multirow{3}{*}{Y$_2$} & x & 3.5 & 0.0 & 0.0 && \\
                       & y & 0.0 & 3.5 & 0.0 &  4.0 & 3.6\\
                       & z & 0.0 & 0.0 & 3.9 && \\
\hline
\multirow{3}{*}{Mn}    & x & 3.1 & 0.2 &-0.4 && \\
                       & y & 0.2 & 3.4 & 0.2 & 4.0 & 3.5 \\
                       & z & 0.2 & -0.1 & 4.0 && \\
\hline
\multirow{3}{*}{O$_1$} & x & -2.0 &  0.1 & 0.3 && \\
                       & y &  0.1 & -1.9 & -0.1 & -2.7 & -2.3 \\
                       & z & -0.1 &  0.1 & -3.1 && \\
\hline
\multirow{3}{*}{O$_2$} & x & -2.0 & 0.2 & 0.2 && \\
                       & y &  0.2 &-1.8 &-0.1  & -2.7 & -2.3 \\
                       & z & -0.1 & 0.1 & -3.3 && \\
\hline
\multirow{3}{*}{O$_3$} & x & -2.9 & 0.0 & 0.0 && \\
                       & y &  0.0 & -2.9 & 0.0 & -2.7 & -2.5 \\
                       & z &  0.0 & 0.0 & -1.6 && \\
\hline
\multirow{3}{*}{O$_4$} & x & -2.9 & -0.2 & 0.0 && \\
                       & y &  0.2 & -2.9 & 0.0 & -2.7 & -2.5 \\
                       & z &  0.0 & 0.0 & -1.5 && \\
\hline
\end{tabular}
\caption{Born effective charge. The dynamic charge, $Tr(Z^{\star})/{3}$,
  is compared to the experimental evaluation from infra-red measurements~\cite{IR-Vinh}.}
\label{t:BCh}
\end{center}
\end{table}

Finally we computed the Born effective charge tensor $Z^{\star}$ for the
optimized geometry. They are reported in table~\ref{t:BCh}. Again our results
compare well with the experimental evaluations validating our methodological
choices.

\section{The phonons spectra}
Table~\ref{t:ph} displays the $\Gamma$ point optical transverse (TO) phonon
modes, both for the paraelectric (columns 1-2) and ferroelectric (columns 6-7)
phases. Raman
and Infra-Red (IR) experimental data are given for comparison. The $P6_3/mmc$
group of the paraelectric phase decomposes in
$$ 1 A_{1g} \oplus 3A_{2u} \oplus 3B_{1g} \oplus 2 B_{2u}\oplus 1E_{1g}  
\oplus 3E_{1u}  \oplus 3E_{2g} \oplus 2E_{2u} $$ 
optical modes, out of which the $A_{1g}$, $E_{1g}$ and
$E_{2g}$ are Raman active while the $A_{2u}$ and $E_{1u}$ are IR active,
leaving the $B_{1g}$, $B_{2u}$ and $E_{2u}$ modes inactive both in Raman and
IR. The $P6_3cm$ group of the ferroelectric phase decomposes in 
$$9A_1 \oplus 5A_2 \oplus 5B_1 \oplus 10B_2 \oplus 14E_1 \oplus  15E_2  $$
optical modes, out of which the $A_1$, $E_1$ are active both in Raman and IR,
the $E_2$ are active in Raman only and the $A_2$, $B_1$ and $B_2$ are inactive
both in Raman and IR.

For the ferroelectric phase, the optical longitudinal (LO) phonon modes have
also been evaluated and are reported in the supplementary data.

\begin{table*} 
  \centering
\begin{tabular}{c@{\;\;}cc@{\;}cc@{\;}c@{\quad}c@{\;\;}c@{\;\;\;}cccccc}
\hline
\multicolumn{4}{c}{This work} &Raman& 
                 \multicolumn{2}{c}{This work}
                 &\multicolumn{2}{c}{IR}&\multicolumn{3}{c}{Raman}\\
 \multicolumn{2}{c}{$P6_3/mmc$} &\multicolumn{2}{c}{$P6_3mc$} &
                 Fukumura~[\onlinecite{Rm_Fuku}] &
                 \multicolumn{2}{c}{$P6_3cm$} & 
 \multicolumn{2}{c}{Zaghrioui~[\onlinecite{IR-Vinh}]}& Iliev~[\onlinecite{Rm_Iliev}]&
Fukumura~[\onlinecite{Rm_Fuku}]&
\multicolumn{2}{c}{Vermette~[\onlinecite{Rm_Vermette}]}\\
Irrep    &TO &Irrep&TO     & 1100\,K & Irrep &TO & 10\,K & 300\,K & 300\,K & 15\,K & 10\,K & 300\,K \\
\hline
         &   &     &   &   & $A_2$ & 33&   &   &       &       &   &    \\
$E_{2u}$  &121&$E_2$&121&120& $E_2$ &119&   &   &    135&    141& - & -  \\
$B_{1g}$  &187&$B$  &179&   & $B_2$ &136&   &   &       &       &   &   \\
         &   &     &   &   & $B_1$ &147&   &   &       &       &   &    \\
$E_{2g}$  &159&$E_2$&148&   & $E_2$ &150&   &   &\bf 190&     - &-  &-  \\
$A_{2u}$  &109&$A$  &138&   & $A_1$ &186&163&154&    148&    160& 161 & 151	\\
$E_{1u}$  &187&$E_1$&177&   & $E_1$ &186&167&162&     - &     - & - & - \\
         &   &     &   &   & $E_1$ &211&211&207&     - &\bf 210& -&-    \\
         &   &     &   &   & $E_2$ &213&   &   &    215&     - & -& - \\
$B_{2u}$  &211&$B$  &204&   & $B_2$ &218&   &   &       &       &  &   \\
         &   &     &   &   & $B_2$ &233&   &   &       &       &  & 	  \\
$E_{2g}$  &257&$E_2$&235&   & $E_2$ &233&   &   &     - &   225 & 231 & 223 \\
$E_{1u}$  &264&$E_1$&241&   & $E_1$ &238&257&249&     - &    247& -&-	 \\
         &   &     &   &   & $A_2$ &241&   &   &       &       &  &     \\
         &   &     &   &   & $A_1$ &262&239&235&     - &     - & 244 & 241  \\
         &   &     &   &   & $E_1$ &269& - & - &     - &     - & -& - \\
         &   &     &   &   & $E_2$ &270&   &   &     - &     - &- &  -\\
         &   &     &   &   & $B_1$ &276&   &   &       &       &  &   \\
         &   &     &   &   & $B_2$ &289&   &   &       &       &  &  \\
         &   &     &   &   & $A_1$ &292&266&260&    257&    264& 264 & 259   \\
         &   &     &   &   & $E_2$ &294&   &   &    302&    307&  &     \\
         &   &     &   &   & $B_1$ &302&   &   &       &       &  &  \\
         &   &     &   &   & $E_1$ &303&301&299&     - &     - & -& -	\\
         &   &     &   &   & $E_1$ &326& - & - &     - &     - & -& - \\%\bf 360&\bf  361 &\bf  354 \\
         &   &     &   &   & $A_2$ &333&   &   &       &       &  &  \\
         &   &     &   &   & $A_1$ &334&311&304&    297&     - & 307 & 300 \\
         &   &     &   &   & $E_2$ &335&   &   &     - &    331&- & -  \\
         &   &     &   &   & $B_2$ &369&   &   &       &       &  &  \\
         &   &     &   &   & $E_2$ &383&   &   &     - &     - & 357 & 356	\\
         &   &     &   &   & $E_1$ &386& - & - &     - &    360&     361 &     354 \\
$E_{2u}$  &402&$E_2$&402&395& $E_2$ &392&   &   &     - &    406& -& -\\
         &   &     &   &   & $E_1$ &404&381&380&    376&    377&- &-	\\
         &   &     &   &   & $B_2$ &430&   &   &       &       &  & 	     \\
         &   &     &   &   & $E_2$ &434&   &   &     - &    444& 441 & 439 \\
         &   &     &   &   & $E_1$ &437&409&400&    408&     - & -&-\\
$E_{1u}$  &426&$E_1$&421&420& $E_1$ &451&423&416&     - &    420&- & - \\
         &   &     &   &   & $E_2$ &454&   &   &     - &     - & -& -\\
         &   &     &   &   & $B_1$ &458&   &   &       &       &  &    \\
$E_{1g}$  &483&$E_1$&471&   & $E_1$ &461& - & - &     - &    509&- & - \\
         &   &     &   &   & $A_2$ &462&   &   &       &       &  &   \\
         &   &     &   &   & $A_1$ &465& - & - &     - &     - & -& -\\
$E_{2g}$  &500&$E_2$&484&   & $E_2$ &473&   &   &     - &    483&- &-\\
$A_{2u}$  &518&$A$  &467&   & $A_1$ &492&434&432&    433&    435& 434 & 431	\\
$B_{1g}$  &495& $B$ &457&   & $B_2$ &504&   &   &       &       &  &  \\
         &   &     &   &   & $E_2$ &515&   &   &     - &     - & -&-  \\
         &   &     &   &   & $E_1$ &515& - & - &     - &     - &- &- \\
         &   &     &   &   & $A_1$ &534&489&486&    459&    466& 467 & 461  \\
         &   &     &   &   & $B_1$ &552&   &   &       &       &  &   \\
         &   &   &&        & $A_2$ &561&   &   &       &       &  &   \\
         &   &   &&        & $B_2$ &567&   &   &       &       &  &   \\
$A_{2u}$  &611&$A$ &600 &   & $A_1$ &612&565&562&     - &     - & -&-	\\
         &   &   &&        & $E_2$ &643&   &   &     - &   647&- & -   \\
         &   &   &&        & $E_1$ &643&594&594&     - &    - & -&-\\
         &   &   &&        & $E_2$ &667&   &   &     - &    - & -&-\\
         &   &   &&        & $E_1$ &668& - & - &    632&   638& 637 & 631 \\
$A_{1g}$  &771&$A$ &735& 666& $A_1$ &722& - & - &    681&   686& 686 & 683  \\
$B_{2u}$  &772&$B$ &729&    & $B_2$ &728&   &   &       &      &  & 	     \\
$B_{1g}$  &868&$B$ &827&    & $B_2$ &813&   &   &       &      &  & 	 \\
\hline 
\end{tabular}
\caption{Hexagonal $\rm YMnO_3$ vibrational frequencies in $\rm cm^{-1}$ for
  the paraelectric ($P6_3/m mc$) and ferroelectric ($P6_{3} cm$) phases. The
  frequencies for one of the possible intermediate phases ($P6_{3} mc$) was
  also computed. Modes with problematic assignment
  are in bold faces~; modes that are symmetry allowed but not seen are designed
  by dashes.}  
\label{t:ph}
\end{table*}

\subsection{High temperature discussion}
At high temperature only four modes were experimentally measured by Fukumura
{\it et al}~\cite{Rm_Fuku}.  These modes were respectively assigned by
Fukumura {\it et al} to one $A_{1g}$ mode ($666\,\rm cm^{-1}$), one $E_{1g}$
mode ($420\,\rm cm^{-1}$) and two $E_{2g}$ modes (120 and $395\,\rm cm^{-1}$)
of the $P6_3/m mc$ group of the high temperature PE phase. In addition
figure~5 of reference~\onlinecite{Rm_Fuku} clearly shows that these modes are
observed continuously from 300\,K to 1200\,K in the same energy ranges. Going
back to our calculations, if we want to assign these modes by continuity from
the low tempera\-ture phase ($P6_3 cm$ to $P6_3/m mc$), we have to suppose
that the mode at $395\,\rm cm^{-1}$ should be assigned as $E_{2u}$ (computed
at $402\,\rm cm^{-1}$ in the $P6_3/m mc$ group and at $392\,\rm cm^{-1}$ in
the $P6_3 cm$ group, and measured at $406\,\rm cm^{-1}$ in the low temperature
phase) and not as $E_{2g}$ (computed at $500\,\rm cm^{-1}$ in the $P6_3/m mc$
group and at $473\,\rm cm^{-1}$ in the $P6_3 cm$ group, and measured at
$483\,\rm cm^{-1}$ in the low temperature phase). Similarly, the mode at $420
\,\rm cm^{-1}$ should be assigned as $E_{1u}$ and not as $E_{1g}$ and the mode
at $120\,\rm cm^{-1}$ should be assigned as $E_{2u}$ rather than $E_{2g}$. At
this point there is a clear contradiction between the continuity requirement
of the phonons modes between the low and high temperature phase and the Raman
symmetry requirements. Indeed the ``ungerade'' $E_{1u}$ and $E_{2u}$
irreducible representations (irreps) are not Raman active and no crystal
disorientation or twinning could induce an inversion between a ``gerade'' and
an ``ungerade'' mode. We must thus assume that the symmetry group at 1200\,K
cannot be the $P6_3/m mc$ group.

What type of hypotheses are left?  The first possibility is the existence of
an intermediate phase in the 1200\,K temperature range as suggested by several
authors~\cite{TE-Abra,TE-Nenert1,TE-Nenert2}. However the $P6_3/mcm$ symmetry
group suggested is no more than the $P6_3/m mc$ group, compatible with the
phonons spectrum. Indeed, symmetry group analysis shows that the $P6_3/m mc$
to $P6_3/mcm$ transformation only exchange the $B_{1g}$ with the $B_{2g}$, and
the $B_{1u}$ with the $B_{2u}$ irreps. %
A second possibility would be that the intermediate phase corresponds to the
$P6_3mc$ group. In this group the inversion center is lost, and the $E_{1g}$
and $E_{1u}$ irreps of $P6_3/m mc$ are associated with the $E_1$ irrep of the
$P6_3mc$ group and with the $E_1$ irrep of the low temperature phase
(resp. $E_{2g}$ and $E_{2u}$ are associated with $E_2$). Both the $E_1$ and
$E_2$ irreps are Raman active in the $P6_3mc$ group and the experimental
Fukumura {\it et al}~\cite{Rm_Fuku} data can be easily associated with the
computed modes (see table~\ref{t:ph}, columns 3-4). The average error between
the computed and measured frequencies is 18\,cm$^{-1}$, essentially supported
by the higher $A$ mode (3\,cm$^{-1}$ for all the other modes).  The main
problem with this hypothesis is that the unit cell tripling at the $P6_3 mc$
to $P6_3cm$ transition should have been seen in diffraction experiments, while
it does not seem to be the case~\cite{TE-Nenert1}. %
Therefore, the only possibility compatible with all data seems to be that a
$P6_3/m mc$ to $P6_3cm$ phase transition occurs at a temperature higher than
1200\,K, as proposed by Jeong~\cite{TE-Jeong} and Gibbs~\cite{TE-Gibbs} from
neutrons diffraction measurements ($T_c=1258\,{\rm K}\pm 14\,\rm K$). In other
terms, all the high temperature data from Fukumura {\it et al}~\cite{Rm_Fuku}
should be interpreted within the ferroelectric $P6_3cm$ group.

%================================================
\subsection{Low temperature discussion}
At low temperature our computed modes fit quite well with the available
experimental data, both Raman and IR (see table~\ref{t:ph}, columns 6 and
further). Table~\ref{t:err} displays the average errors between our computed
modes and the experimental data for each irreducible representation.
 \begin{table} 
  \centering
\begin{tabular}{c@{\;\;}cc@{\quad}cc@{\;\;}c@{\;\;}c}
\hline
Irrep & \multicolumn{2}{c}{IR} & \multicolumn{4}{c}{Raman} \\
& \multicolumn{2}{c}{Zaghrioui~[\onlinecite{IR-Vinh}]}& Iliev~[\onlinecite{Rm_Iliev}]&
Fukumura~[\onlinecite{Rm_Fuku}]&
\multicolumn{2}{c}{Vermette~[\onlinecite{Rm_Vermette}]}\\
 & 10\,K & 300\,K & 300\,K & 15\,K & 10\,K & 300\,K \\
\hline 
$A_1$ & 14 & 16 & 20 & 21 & 15 & 17\\
$E_1$ & 8  & 8  & 18 & 11 & 19 & 24\\
$E_2$ &    &    & 11 & 4  & 9  & 10 \\
\hline
\end{tabular}
\caption{Average errors per irrep on the 
  $\rm YMnO_3$ vibrational frequencies compared to different experimental data
  (in $\rm cm^{-1}$).}  
\label{t:err}
\end{table}
The experimental phonons modes can be associated with the computed ones except
for a few exceptions. %
i) The mode at 190\,cm$^{-1}$ seen by Iliev {\it et al} in the $A_1$
irreducible representation cannot be associated with a computed mode. This
mode, not seen by the other authors, cannot be associated with a $A_1$
mode. Indeed, the $A_1$ modes are IR active and all predicted modes are
observed by Zaghrioui {\it et al}~\cite{IR-Vinh} up to more than 450 wave
numbers. If not artifactual, the 190\,cm$^{-1}$ mode must belong to another
irreducible representation. The only possibility in this energy range is an
$E_2$ mode predicted at 150 cm$^{-1}$. Indeed, all other IR or Raman active
modes in this energy range are experimentally observed and easily associated
with the computed phonons. %
ii) The  210\,cm$^{-1}$ $A_1$ mode seen by Fukumura {\it et
  al}~\cite{Rm_Fuku}. Once again a $A_1$ mode around  210\,cm$^{-1}$ is
incompatible both with the computed values and with the other experimental
results (IR or Raman). Most probably this mode has been improperly assigned as
a $A_1$ mode and should rather be considered as a $E_1$ (seen at
211\,cm$^{-1}$ by Zaghrioui {\it et al}~\cite{IR-Vinh}) or a $E_2$ mode (seen
at 215\,cm$^{-1}$ by  Iliev {\it et al}~\cite{Rm_Iliev}).  

Another point we would like to address is the reason why all phonons
modes are not observed in experimental data. As far as the IR active modes are
concerned we computed an estimate of the intensities, the analysis of which
gives us some insight on the reasons some modes are not observed. 
In the $A_1$ representation two modes are not seen in
reference~\onlinecite{IR-Vinh}, namely the highest mode computed around
720\,cm$^{-1}$ and the mode computed around 460\,cm$^{-1}$.  The IR computed
intensity of the 720\,cm$^{-1}$ $A_1$ mode is very weak (between 5 and 12
km/mol according to the functional used) and it is thus not surprising this
mode is not seen experimentally. The IR intensity of the $A_1$ mode computed
around 460\,cm$^{-1}$ mode is of the same order of magnitude as the next
computed $A_1$ mode at 492\,cm$^{-1}$ which was associated to the experimental
mode found at 434\,cm$^{-1}$). Even if the IR intensity remains small it is non
negligible and the second mode is observed. Assuming
an equivalent shift in energy between experiments and calculation on
these two modes, the first mode should be searched around 400\,cm$^{-1}$ in
the IR spectrum  however it is not seen. 
In the $E_1$ irreducible representation, five modes are not seen in IR
experiments, namely the modes computed at 269\,cm$^{-1}$, 386\,cm$^{-1}$,
461\,cm$^{-1}$, 515\,cm$^{-1}$ and 668\,cm$^{-1}$. The calculations reveals
that all these modes are associated with very low IR intensities that could
explain there are not seen by Zaghrioui {\it et al}~\cite{IR-Vinh}.

\subsection{The $A_2$ phonon mode at 33\,cm$^{-1}$}
\begin{figure}[h] 
\resizebox{9cm}{!}{\includegraphics{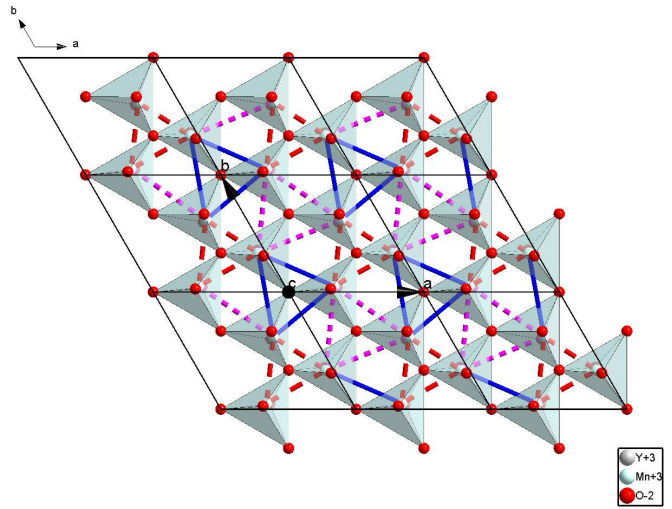}}
\caption{(Color online) Schematic picture of the atomic displacements in the
  $(\vec a,\vec b)$ plane for the lowest $A_2$ phonon mode. The blue, solid
  lines, Mn triangles remain nearly untouched and rotate as a whole around the
  $\vec c$ axis. The red, dashed and magenta, dotted Mn triangles are
  alternatively enlarged/reduced, the Mn ions moving within the $(\vec a,\vec
  b)$ plane while the central oxygens move conversely along the $\vec c$
  direction. A jmol xyz file can be found in the supplementary data in order
  to visualize the displacements of one manganese plane.}
\label{fig:A2}
\end{figure}

We would now like to discuss in a little more details the first optical mode,
computed at $33\rm\,cm^{-1}\simeq 4\,meV$ in the $A_2$ irreducible
representation.  The displacements vector associated with this phonon mode is
reported on table~\ref{t:A2} and pictured in figure~\ref{fig:A2}.
\begin{table}[h]
  \centering
  \begin{tabular}{c@{\quad}ccc@{\qquad}ccc}
\hline
Atom & \multicolumn{3}{c}{Positions} & \multicolumn{3}{c}{Displacements} \\
     &  x & y & z & x & y & z \\
\hline
$\rm Y_1$ &  0.0000 &  0.0000 & $z_{Y_1}$ &  0.0000 &  0.0000 &  0.0000  \\
$\rm Y_2$ &  $1/3$  &  $-1/3$ & $z_{Y_2}$ &  0.0000 &  0.0000 & -0.0713  \\
$\rm Mn_1$&$x_{Mn_1}$ &  0.0000 & $z_{Mn_1}$ &  0.0322 &  0.0558 &  0.0000  \\ 
% MN & -0.0322    & -0.0558    &  0.0000  \\ 
% MN & -0.0645    &  0.0000    &  0.0000  \\ 
% MN &  0.0322    & -0.0558    &  0.0000  \\ 
% MN & -0.0322    &  0.0558    &  0.0000  \\ 
% MN &  0.0645    &  0.0000    &  0.0000  \\ 
$\rm O_1$ & $x_{O_1}$ &  0.0000 & $z_{O_1}$ &  0.0206 &  0.0357 &  0.0000  \\ 
% O  & -0.0206    & -0.0357    &  0.0000  \\ 
% O  & -0.0412    &  0.0000    &  0.0000  \\ 
% O  &  0.0206    & -0.0357    &  0.0000  \\ 
% O  & -0.0206    &  0.0357    &  0.0000  \\ 
% O  &  0.0412    &  0.0000    &  0.0000  \\ 
$\rm O_2$ &$x_{O_2}$ &  0.0000 & $z_{O_2}$ &  0.0272 &  0.0471 &  0.0000  \\ 
% O  & -0.0272    & -0.0471    &  0.0000  \\ 
% O  & -0.0544    &  0.0000    &  0.0000  \\ 
% O  &  0.0272    & -0.0471    &  0.0000  \\ 
% O  & -0.0272    &  0.0471    &  0.0000  \\ 
% O  &  0.0544    &  0.0000    &  0.0000  \\ 
$\rm O_3$ & 0.0000 &  0.0000 & $z_{O_3}$ &  0.0000 &  0.0000 &  0.0000  \\ 
% O  &  0.0000    &  0.0000    &  0.0000  \\ 
$\rm O_4$ & $1/3$  &  $2/3$  & $z_{O_4}$ &  0.0000 &  0.0000 & -0.0145  \\ 
% O  &  0.0000    &  0.0000    & -0.0145  \\ 
% O  &  0.0000    &  0.0000    &  0.0145  \\ 
% O  &  0.0000    &  0.0000    &  0.0145  \\ 
\hline
  \end{tabular}
  \caption{Displacements vector (normalized to classical amplitudes) 
associated with the first $A_2$ mode found  around 33\,cm$^{-1}$. }
  \label{t:A2}
\end{table}

This phonon mode, active neither in Infra-Red nor in Raman experiments
exhibits displacements that can be expected to strongly affect the magnetic
exchange coupling constants along the red, dashed and magenta, dotted bonds as
picture in figure~\ref{fig:A2}.

\begin{figure}[h]
\begin{minipage}{4cm}
  ($J_{l_1}$)\resizebox{3.2cm}{!}{\includegraphics{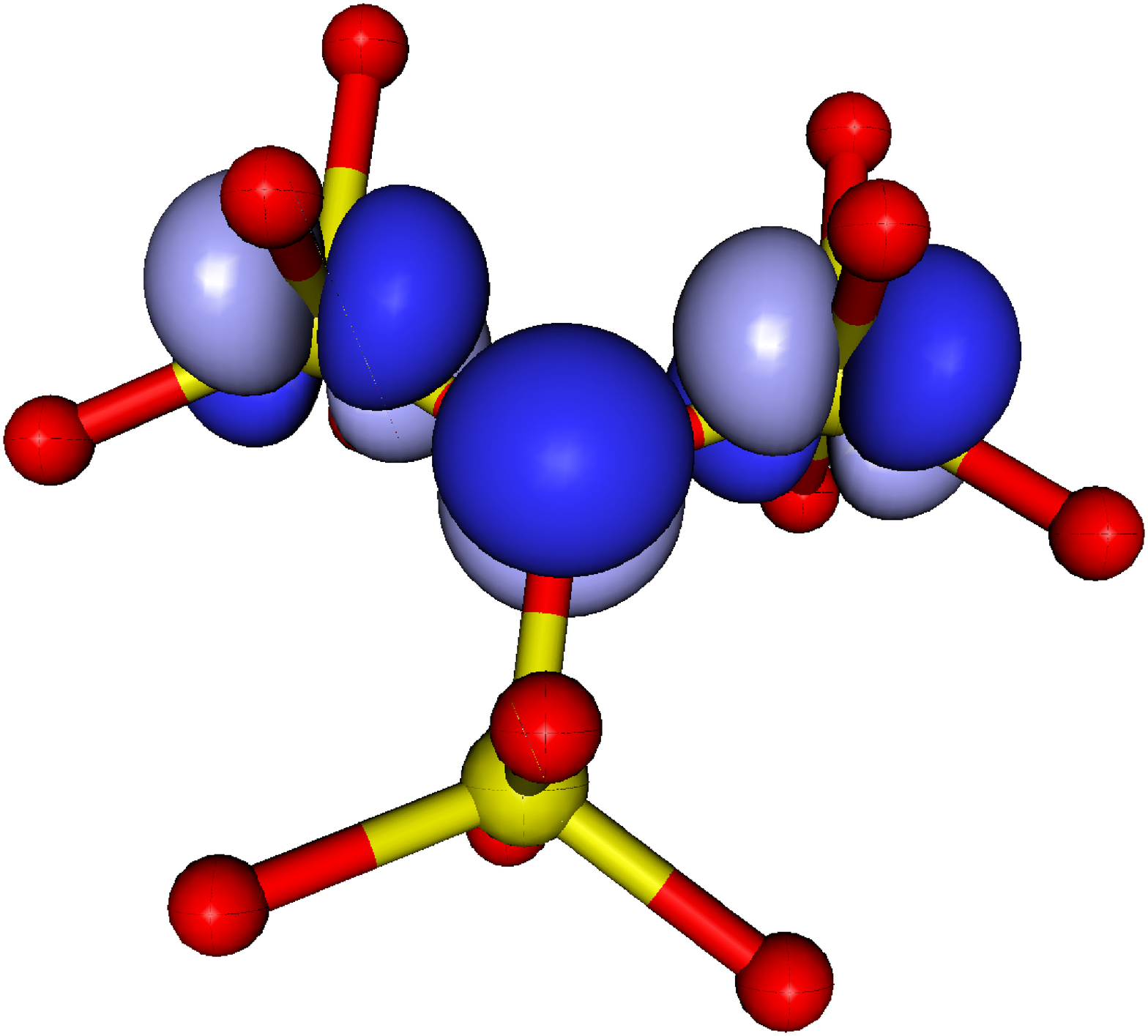}} \\
  ($J_{l_3}$)\resizebox{3.2cm}{!}{\includegraphics{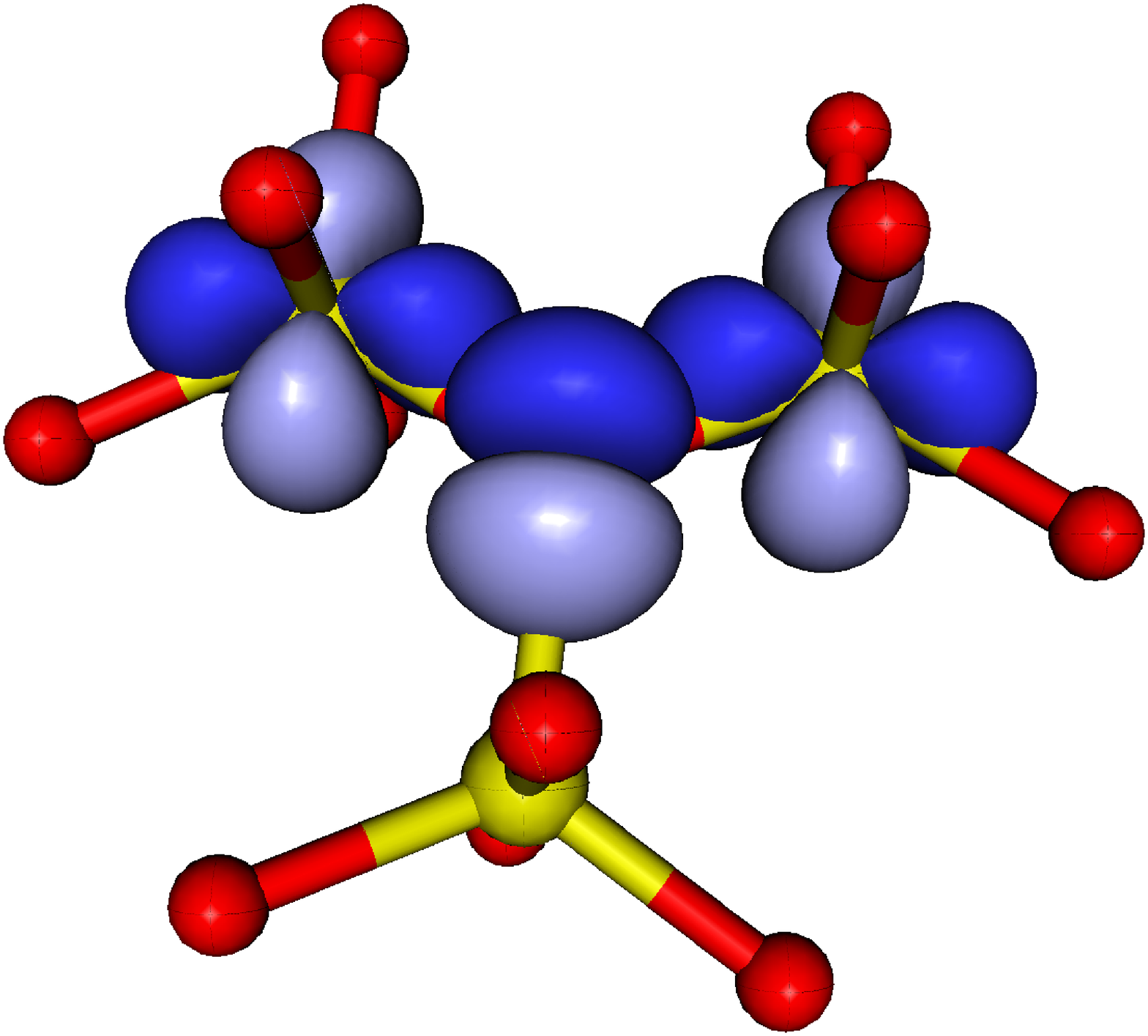}} 
\end{minipage}
\begin{minipage}{4cm}
  ($J_{l_2}$)\resizebox{3.2cm}{!}{\includegraphics{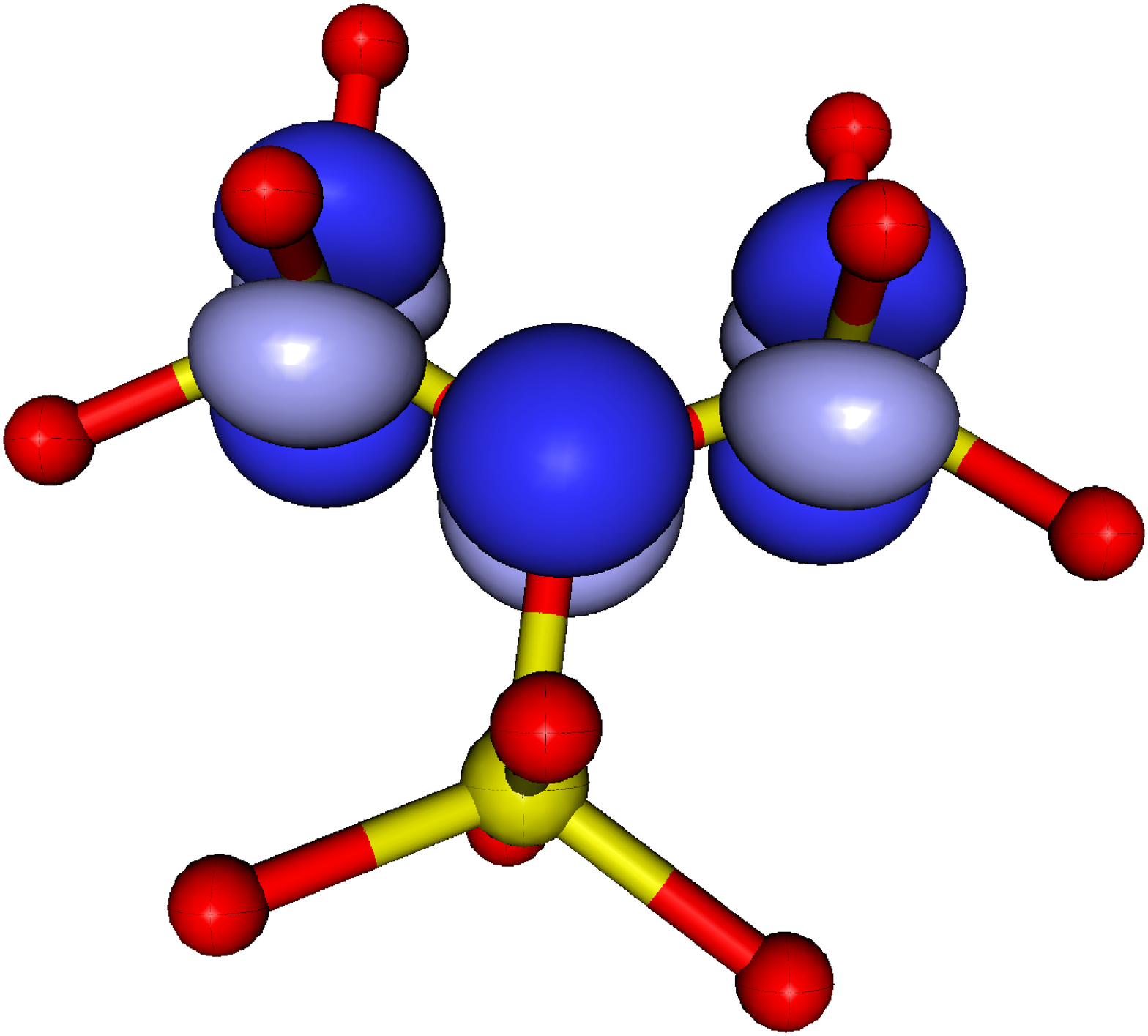}} \\
  ($J_{l_4}$)\resizebox{3.2cm}{!}{\includegraphics{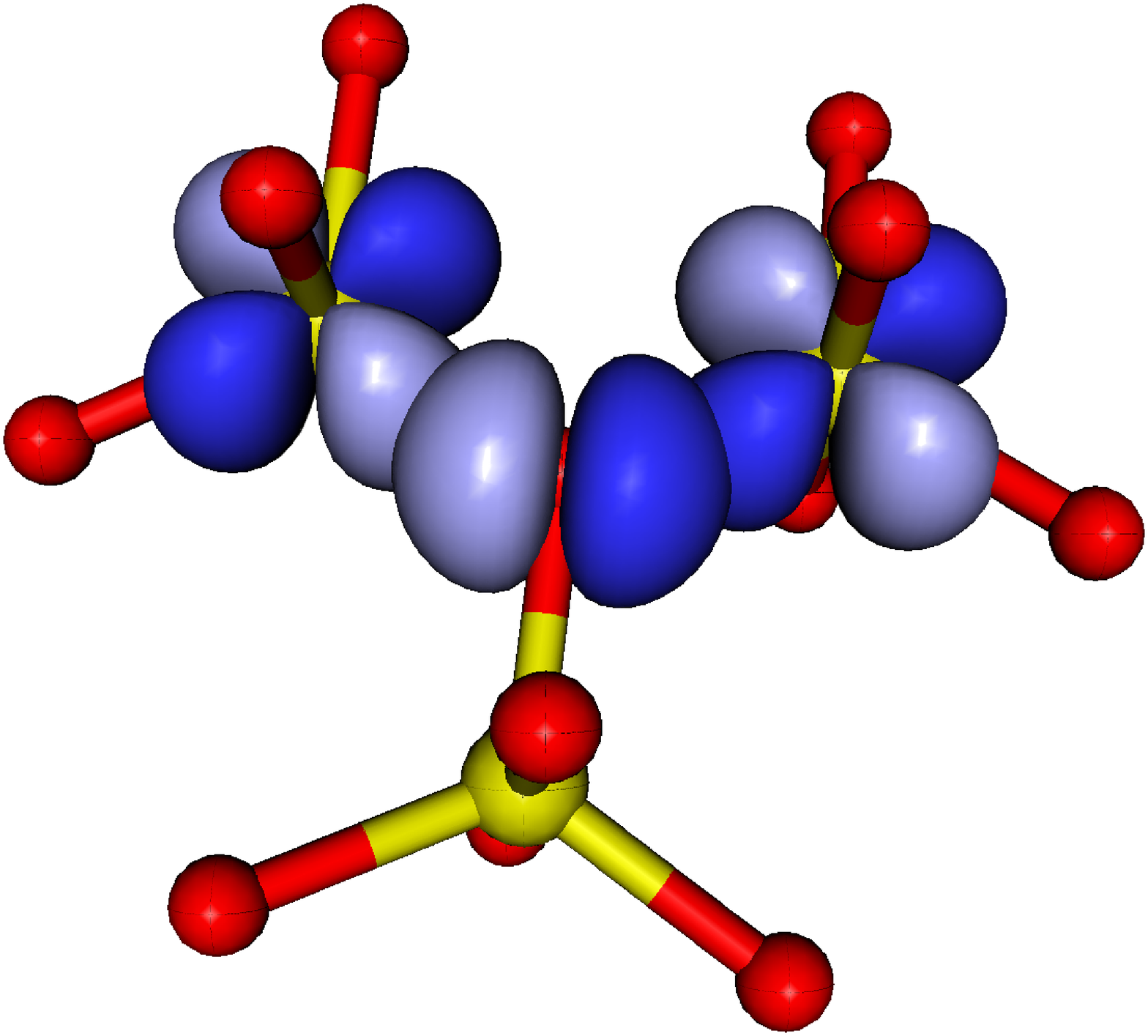}} 
\end{minipage}
  \caption{Main super-exchange paths between two manganese atoms. Let us
    remember that in $\rm YMnO_3$ the manganese is $3d^4$ and the singly 
    occupied orbitals are $d_{xy}$, $d_{x^2-y^2}$, $d_{xz}$, $d_{yz}$ if $z$
    is along the $\vec c$ axis. We can evaluate\\
    $J_{l_1} \sim \langle {\rm Mn}_1 d_{xz} | {\rm O} p_z\rangle^2\,
               \langle {\rm O} p_z | {\rm Mn}_2 d_{xz}\rangle^2$ \\
    $J_{l_2} \sim \langle {\rm Mn}_1 d_{yz} | {\rm O} p_z\rangle^2\,
               \langle {\rm O} p_z | {\rm Mn}_2 d_{yz}\rangle^2$ \\
    $J_{l_3} \sim \langle {\rm Mn}_1 d_{x^2-y^2} | {\rm O} p_y\rangle^2\,
               \langle {\rm O} p_y | {\rm Mn}_2 d_{x^2-y^2}\rangle^2$ \\
   $J_{l_4} \sim \langle {\rm Mn}_1 d_{xy} | {\rm O} p_x\rangle^2\,
               \langle {\rm O} p_x | {\rm Mn}_2 d_{xy}\rangle^2$ 
             }
  \label{fig:coupl}
\end{figure}

Let us notice that the effective exchange $J$ between two
manganese atoms is the sum of a direct exchange contribution $J_d$ (Pauli
exchange, always ferromagnetic, exponentially dependent on the Mn-Mn distance)
and a through-ligand super-exchange term $J_l$ (antiferromagnetic, dependent
on the metal ligand distances and on the Mn-O-Mn angles). In $\rm YMnO_3$ the
magnetic exchange couplings are known to be antiferromagnetic showing that the
through-ligand super-exchange term $J_l$ dominates over the direct exchange
contribution $J_d$. We analyzed this super-exchange term and reported in
figure~\ref{fig:coupl} the main super-exchange paths. $J_l$ thus scales as the sum
of the four associated super-exchange terms, each of them scaling as the 
square of the product of the associated ${\rm Mn}_1 3d - {\rm O} 2p$
orbital overlap and the  ${\rm O} 2p - {\rm Mn}_2\, 3d$ orbital overlap.

The $A_2$ phonon mode under consideration involves the following atomic
displacements on the red, dashed and magenta, dotted Mn triangles (see
figure~\ref{fig:A2}).
\begin{itemize}
\item The Mn-Mn bond length elongation\,/\,contraction (in opposite phases
  between red and magenta triangles). This movement affects the direct part of
  the exchange~: $J_d$, increasing its ferromagnetic character when the Mn-Mn
  distances are contracted and decreasing it when the Mn-Mn distances are
  elongated.
\item The up and down movement (along the $c$ direction) of the oxygens atoms
  at the center of these triangles. This movement strongly affects both the
  Mn-O distances, reducing them when the oxygens come closer to the Mn plane,
  and the Mn-O-Mn angle. 
\end{itemize}
One sees immediately using figure~\ref{fig:coupl} that, when the oxygen comes
closer to the Mn plane, the overlap between the magnetic orbitals of the Mn
atoms and the $2p$ bridging orbitals of the central oxygen increases, thus
increasing the antiferromagnetic character of $J_l$. At the same time the
Mn-Mn bond length increases thus decreasing the ferromagnetic character of
$J_d$. Both terms goes in the same direction resulting in a enlarged
antiferromagnetic character for the bonds of the (alternatively red/magenta)
triangles when the central oxygen comes closer to the Mn plane and a reduced
antiferromagnetic character when the central oxygen goes away from the Mn
plane.%

The blue, solid line, Mn triangles (centered around the unit cell origin) are
subject to a rotation {\em as a whole} around the $\vec c$ axis going through
their central oxygen ($\rm O_3$). One thus does not expect the magnetic
exchange within the blue Mn triangles to be affected by this $A_2$ phonon
mode. However this rotation results in a strong distortion of the triangles
with blue red and magenta bond inducing a magnetic interaction scheme quite
far from the nearly homogeneous triangular lattice of the static structure.
 
The main conclusion of this analysis is that this $A_2$ phonon mode is
strongly coupled to the magnetic interactions. 
 
Let us now put these results into perspective with the characteristics of the
hybrid phonon-magnon excitation, studied using inelastic polarized and
non-polarized neutron scattering by Pailh\`es and coworkers~\cite{Pailhes09}.
One notices that this $A_2$ mode seems to fulfill all the requirements for a
good candidate for the phonon participating to the hybrid mode. Indeed, it is
an optical mode in the adequate energy range, exhibiting atomic displacements
consistent with the umbrella motion proposed in
reference~\onlinecite{Pailhes09} and that can be expected from the above
analysis to be strongly coupled to the spin degree of freedom.

\section{Conclusion}
The present paper proposes $\Gamma$ point phonon calculations for the $\rm
YMnO_3$ compound both in the ferroelectric phase and in the paraelectric
phase.  Our calculations agrees well with the Infra-Red and Raman experimental
data. The experimental versus calculated modes correspondence is discussed in
details 
for the few problematic modes. We were able to explain the fact that several
modes could not be seen in IR experiments due to their very low (calculated)
intensities. %
A careful analysis of the phonons modes
correspondence, when going from the ferroelectric to the paraelectric
phase, is performed and leads to the conclusion that, the phonon modes
observed in Raman scattering at 1200K~\cite{Rm_Fuku} cannot be associated with
the paraelectric phase $P6_3/m mc$ space group. The different intermediate
subgroups allowed between the paraelectric $P6_3/m mc$ space group and the
ferroelectric $P6_{3} cm$ group were also checked against the modes continuity
and the experimental data and had to be discarded. The only possibility
agreeing with the phonon calculations and the experimental data is that the
$P6_3/m mc$ to $P6_3cm$ phase transition occurs at a temperature higher than
1200\,K, supporting the proposition of Jeong~\cite{TE-Jeong} and
Gibbs~\cite{TE-Gibbs} ($T_c=1258\,{\rm K}\pm 14\,\rm K$).

Finally we studied in more details the first optical mode. This mode belongs
to $A_2$ irreducible representation and is inactive both in IR and Raman
experiment. This mode however seems to have all necessary characteristics to
be a good candidate for the phonon part of the electro-magnon mode observed in
inelastic neutrons scattering by Petit, Pailh\`es {\it et
  al}~\cite{Petit07,Pailhes09}.

%=========================================================================
\acknowledgments The authors thank V. TaPhuoc and Ph. Ghosez and collaborators
for helpful discussions. This work was done with the support of the French
national computer center IDRIS under project n$\circ$ 081842 and the regional
computer center CRIHAN under project n$\circ$ 2007013.

%=========================================================================


\begin{thebibliography}{9999}

% Electro-magnons en général - prédiction de l'existence 
\bibitem{EM} V. G. Baryaktar and I. E. Chapius, Sov. Phys. Solid State {\bf 11},
2628 (1970)~; I. A. Akhiezer and L. N. Davydov, Sov. Phys. Solid State {\bf 12},
2563 (1971).

% Electro-magnons dans autres RMnO3 / hexagonaux
\bibitem{EM-RMnO3} A. Pimenov, A. A. Mukhin, V. YU. Ivanov, V. D. Travkin,
  A. M. Balbashov and A. Loidl, Nature Physics {\bf 2}, 97 (2006). 

% Electro-magnons dans YMnO3
\bibitem{Petit07} S. Petit, F. Moussa, M. Hennion, S. Pailhe\`s,
  L. Pinsard-Gaudart and A. Ivanov, Phys. Rev. Letters {\bf 99}, 266604 (2007).
%\bibitem{Petit08} S. Petit, S. Pailh\`es, X. Fabr\`eges, M. Hennion, F. Moussa, L. Pinsard,
%L.-P. Regnault and A. Ivanov, PRAMANA {\bf 71}, 869 (2008).
\bibitem{Pailhes09} S. Pailh\`es, X. Fabr\`eges, L. P. R\'egnault,
  L. Pinsard-Godart, I. Mirebeau, F. Moussa, M. Hennion and S. Petit,
  Phys. Rev.  {\bf B 79}, 134409 (2009). 

% Structure TFE 
\bibitem{TE=920K} G.~A. Smolenskii and I.~E. Chupis, Sov. Phys. Usp. {\bf 25},
  475 (1982).
\bibitem{TE=1100K} T. Katsufuji, M. Masaki, A. Machida, M. Moritomo, K. Kato,
  E. Nishibori, M. Takata, M. Sakata, K. Ohoyama, K. Kitazawa and H. Takagi,
  Phys. Rev. {\bf B 66}, 134434 (2002).
\bibitem{TE-Abra}  S. Abrahams, Acta Crys. {\bf B 65}, 450 (2009).
\bibitem{TE-Gibbs} A.~S. Gibbs, K.~S. Knight and Ph. Lightfoot,
  Phys. Rev. {\bf B 83}, 094111 (2011).

% Ordre des spins
\bibitem{spins} E.~F. Bertaut, R. Pauthenet and M. Mercier, Physics Letters
  {\bf 7}, 110 (1963).

% TN + ordre des spins 
\bibitem{TN-sp} A. Mu\~noz, J.~A. Alonso, M.~J. Mart\'{i}nez-Lope,
  M.~T. Cas\'ais, and J.~L. Mart\'{i}nez and M.~T. Fern\'andez-D\'{i}az,
  Phys. Rev. {\bf B 62}, 9498 (2000).

% Couplage magnéto-élastique
\bibitem{Park08} S. Lee, A. Pirogov, M. Kang, K.-H. Jang, M. Yonemura,
  T. Kamiyama, S.-W. Cheong, F. Gozzo, N. Shin, H. Kimura, Y. Noda and
  J.-G. Park, Nature {\bf 451} 805, (2008).

% Papier avec Charles 
\bibitem{YMnO3_1} K. Singh, N. Bellido, Ch. Simon, J. Varignon and
  M.-B. Lepetit, to be published elsewhere.  

% Raman 
\bibitem{Rm_Iliev} M. N. Iliev, H.-G. Lee, V. N. Popov, M. V. Abrashev,
  A. Hamed, R. L. Meng, and C. W. Chu, Phys. Rev.  {\bf B 56}, 2488 (1997).
\bibitem{Rm_Fuku} H. Fukumura, S. Matsui, H. Harima, K. Kisoda, T. Takahashi,
  T. Yoshimura and N. Fujimura, J. Phys.: Condens. Matter {\bf 19} 365239
  (2007).
\bibitem{Rm_Vermette} J. Vermette, S. Jandl, A.~A. Mukhin, V. Yu Ivanov,
  A. Balbashov, M.~M. Gospodinov and L. Pinsard-Gaudart, J. Phys.:
  Condens. Matter {\bf 22}, 356002 (2010).

% Phonons IR 
\bibitem{IR-Vinh} M. Zaghrioui, V. Ta Phuoc, R. A. Souza, and M. Gervais,
  Phys. Rev. {\bf B 78}, 184305 (2008).

% CRYSTAL 09
\bibitem{CRYSTAL09} R. Dovesi, R. Orlando, B. Civalleri, C. Roetti,
  V.R. Saunders, C.M. Zicovich-Wilson, Z. Kristallogr. {\bf 220}, 571 (2005)~;
  R. Dovesi, V.R. Saunders, C. Roetti, R. Orlando, C. M. Zicovich-Wilson,
  F. Pascale, B. Civalleri, K. Doll, N.M. Harrison, I.J. Bush, Ph. D’Arco,
  M. Llunell, {\it CRYSTAL09 User’s Manual}, University of Torino, Torino, (2009).

% Fonctionnelles
\bibitem{B3LYP}A.~D. Becke, Phys. Rev. A, {\bf 38}, 3098 (1988).
\bibitem{B1WC}D.~I. Bilc, R. Orlando, R. Shaltaf, G. M. Rignanese,
  J. I{\~n}iguez and P. Ghosez, Phys. Rev. B, {\bf 77}, 165107 (2008).
\bibitem{Perdrew_PRB33} J.~P. Perdew and Y. Wang, Phys. Rev. B, {\bf 33}, 8800
  (1986).

% Bases et pseudos
\bibitem{Bases} Mn and Y~: P.~J. Hay and W.~R. Wadt, J. Chem. Phys. {\bf 82},
  299 (1985); Evarestov et al., Solid State Commun. {\bf 127}, 367 (2003). \\
O~: A. Gell\'e and C. calzado, private communication. 

% Structures
\bibitem{VAken} B.~B. van Aken, A. Meetsma and Th.~T.~M. Palstra, Acta
  Cryst. {\bf C 57}, 230 (2001).

% Polarisation 
\bibitem{VAkenPol} B.~B. van Aken, T.~T.~M. Palstra, A. Filippetti and
  N.~A. Spaldin, Nature Materials {\bf 3}, 164 (2004).
\bibitem{Pol300} S.~H. Kim, S.~H. Lee, T.~H. Kim, T. Zyung, Y.~H. Jeong and
  M.~S. Jang, Crys. Res. Tech. {\bf 35}, 19 (2000).

% Structure TFE suite
\bibitem{TE-Nenert1} G. N\'enert, Y. Ren, H.~T. Stokes and T.~T. M. Palstra,
  arXiv:cond-mat/0504546. 
\bibitem{TE-Nenert2} G. N\'enert, M. Pollet, S. Marinel, G.~R. Blake,
  A. Meetsma and T.~T. M. Palstra, J. Phys.: Condens. Matter {\bf 19}, 466212
  (2007).
\bibitem{TE-Jeong} Il-Kyoung Jeong, N. Hurb and Th. Proffen,
  J. Appl. Cryst. {\bf 40}, 730 (2007).

\end{thebibliography}
\end{document}